\documentclass[12pt]{article}

\usepackage{amsmath}
\usepackage{amsfonts}
\usepackage{latexsym}
\usepackage{mathrsfs}
\usepackage{graphicx}

\newcommand{\tr}{{\rm tr }}

\newcommand{\N}{{\mathscr N}}
\newcommand{\C}{{\mathscr C}}

\newcommand{\F}{{\mathscr F}}

\def\S{{\mathscr S}}

\def\pr{\partial}

\newcommand{\Geff}{\Gamma_{\rm eff}}
\newcommand{\Gcl}{\Gamma_{\rm cl}}

\newcommand{\dv}{\!\!{\rm d}^3x\,}
\newcommand{\dz}{\!\!{\rm d}^4z\,}
\newcommand{\dt}{\!\!{\rm d}t\,}

\newcommand{\al}{\alpha}
\def\da{{\dot\alpha}}

\newcommand{\eps}{\varepsilon}

\newcommand{\Dslash}{\,/\!\!\!\!D}
\newcommand{\dslash}{\,/\!\!\!\partial}

\newcommand{\tfr}[2]{{\textstyle \frac{#1}{#2}}}

\newcommand{\fdq}[2]{\frac{\delta #1}{\delta #2}}
\newcommand{\cdq}[1]{#1 \fdq{}{#1}}

\newcommand{\pdq}[2]{\frac{\partial #1}{\partial #2}}

\renewcommand{\i}{{\rm i}}

\newcommand{\op}{{\rm Op}}
\newcommand{\brs}{{\rm BRS}}
\newcommand{\T}{{\rm T}}

\newcommand{\out}{{\rm out}}
\renewcommand{\in}{{\rm in }}

\begin{document}

\thispagestyle{empty}
\vspace*{-5mm}
\begin{flushright}
hep-th/0101165\\  December 2000
\end{flushright}

\vspace{0.5cm}
\begin{center}
{\Large \bf \!\!\! Supersymmetry Transformation of Quantum Fields II:\!\!\! \\[1.5ex]
Supersymmetric Yang-Mills Theory}

\vspace{0.5cm}

Christian Rupp \footnote{Email: rupp@itp.unibe.ch}\\
Institut f{\"u}r Theoretische Physik\\
Universit{\"a}t Bern\\
Sidlerstrasse 5\\
CH - 3012 Bern\\
Switzerland

Klaus Sibold \footnote{Email: Klaus.Sibold@itp.uni-leipzig.de}\\
Institut f{\"u}r Theoretische Physik\\
Universit{\"a}t Leipzig\\
Augustusplatz 10/11\\
D - 04109 Leipzig\\
Germany
\end{center}

\vspace{1cm}

\begin{center}
\parbox{12cm}{
\centerline{\small \bf Abstract} 
{ \small \noindent 
We study the transformation law of quantum fields in super Yang-Mills
theory quantized in the
Wess-Zumino gauge. It can be derived from a local version of
generalized Slavnov-Taylor identities for general Green
functions. Under suitable normalization conditions the transformations
are local. 
Within the vector multiplet anomalous dimensions become equal.
The breaking of susy shows up in Fock but not in Hilbert space and is
not reflected in the transformation law for the fields.
 }}
\end{center}

\vspace*{10mm}
\begin{tabbing}
\> Anomalous Dimension.
\end{tabbing}
\newpage

\section{Introduction}
The quantization of $N=1$ supersymmetric Yang-Mills theory (SYM) can
be performed in essentially two ways: one is the linear realization of
supersymmetry in terms of superfields, the other one uses the
elimination of longitudinal and auxiliary fields and leads to
non-linear realization of supersymmetry (the Wess-Zumino gauge).
In the linear realization one is faced with an off-shell infrared
problem since the vector superfield is dimensionless.
To overcome it one breaks susy explicitly and can guarantee existence
and susy covariance only for Green functions of gauge (i.e.\ BRS)
invariant operators \cite{PS}.
Similarly in the non-linear version:
at most Green functions of BRS invariant quantities possess proper
covariance under susy because gauge fixing terms usually break
supersymmetry.
Although the treatment of the renormalization procedure in terms of the
vertex functional seems to be in good shape \cite{White, MPW, PW, HKS,
  HKS2} it is not clear how a supersymmetry charge can be defined and
how it acts on the fields. We investigated this question in the
context of the Wess-Zumino model without auxiliary fields and in
supersymmetric QED (in the Wess-Zumino gauge) \cite{RSS}. 
It turned out that the Wess-Zumino model admits a susy charge which
generates the non-linear field transformation and commutes with the
S-matrix i.e.\ is conserved.
In SQED however the susy charge $Q_\al$ becomes time-dependent
according to
\begin{equation}
Q_\al^{\rm out} -  Q_\al^{\rm in} = \left[ Q^\brs, \pdq{\Geff}{\eps^\al}\right]
\end{equation}
where
$Q^\brs$ denotes the BRS charge and
$\partial\Geff/\partial\eps$ arises from the gauge fixing term.
The field transformations are local, but for $\lambda$, $\bar \lambda$
(photino) not consistent with time evolution. It turns out however
that the combinations $\lambda-\eps \bar c$, $\bar\lambda-\bar
\eps \bar c$ -- here $\bar c$ is the antighost field --
indeed have a transformation law which is consistent with time
evolution.

In the present paper we study the analogous situation for SYM where
the complication arises that $B$ -- the Lagrange multiplier field
showing up in gauge fixing -- is no longer a free field. This
difficulty can be overcome by a suitable choice of normalization
conditions, as will be explained in sect. \ref{sec:three}.
This normalization affects the anomalous dimension of $\lambda$ ($\bar
\lambda$) such that it becomes equal to that of the vector.

\section{The model and its vertex functional
  \label{sec:two}}
\setcounter{equation}{0}
We work with SYM, assume a simple gauge group and matter multiplets in
some representation. In order to make the paper self-contained we
write down explicitly the field transformations in the classical approximation
\begin{align}
\label{BRSclassical}
sA{}_\mu & =  \partial_\mu c -ig [c,A_\mu]
             + \i\eps\sigma_\mu \bar\lambda
             -\i \lambda\sigma_\mu\bar \eps
   -\i\omega^\nu\partial_\nu A_\mu
\ ,\\
s\lambda^\alpha & =  -\i g \{c,\lambda^\alpha\}
             +\tfr{\i}{2} (\eps\sigma^{\rho\sigma})^\alpha
             F_{\rho\sigma} + \i\eps^\alpha\, D
             -\i\omega^\nu\partial_\nu  \lambda^\alpha
\ ,\\
s\bar\lambda{}_\da & =  -\i g\{c,\bar\lambda_\da\}
             - \tfr{\i}{2} (\bar\eps\bar\sigma^{\rho\sigma})
             _\da F_{\rho\sigma} + \i\bar\eps_\da\, D 
             -\i\omega^\nu\partial_\nu \bar\lambda_\da
\ ,\\
s\phi_i & =  -\i g c\,\phi_i +\sqrt{2}\, \eps\psi_i - \i\omega^\nu\partial_\nu \phi_i
\ ,\\
s\phi_i^\dagger & =  +\i g (\phi^\dagger c)_i +\sqrt{2}\,
            \bar\psi_i\bar \eps - \i\omega^\nu\partial_\nu \phi_i^\dagger
\ ,\\
s\psi_i^\alpha & =  -\i g c\,\psi_i^\alpha + \sqrt{2}\, 
         \eps^\alpha\, F_i
         -\sqrt{2}\, \i (\bar\eps \bar\sigma^\mu)^\alpha D_\mu\phi_i 
         -\i\omega^\nu\partial_\nu \psi_i^\alpha
\ ,\\
s \bar\psi_{i\da} & =  -\i g ({\bar\psi}_\da c)_i 
         - \sqrt{2}\,\bar\eps_\da\, F^\dagger_i 
         + \sqrt{2}\, \i(\eps\sigma^\mu)_\da (D_\mu\phi_i)^\dagger 
         -\i\omega^\nu\partial_\nu \bar\psi_{i\da}\\
s c & =  -\i gc^2 + 2\i\eps\sigma^\nu\bar\eps A_\nu
         -\i\omega^\nu\partial_\nu c
\ ,\\
s\eps^\alpha & =  0
\ ,\\
s\bar\eps^\da & = 0
\ ,\\
s\omega^\nu & =  2\eps\sigma^\nu\bar\eps
\ .
\end{align}
(We use the conventions of \cite{HKS2}.) The auxiliary fields $D$ and
$F_i$ have to be replaced by the equations of motion. 
The ordinary transformation parameters have been promoted to constant
ghost fields ($\eps$, $\bar\eps$ susy; $\omega^\nu$ translations).

The classical action
\begin{align}
\Gcl &= \Gamma_{\rm inv} + \Gamma_{\rm g.f.}\\
\Gamma_{\rm inv} &= \int \tr \left( -\tfr{1}{4} F^{\mu\nu}F_{\mu\nu} +
  \i \lambda \Dslash \bar\lambda \right) + {\rm matter}\\
\Gamma_{\rm g.f.} &= s \, \int \tr \left( \tfr{\xi}{2} \bar c B + \bar
  c \partial A \right) \label{Gammagf}
\end{align}
is invariant under the generalized BRS transformations.
In order to deal with the non-linearity of the transformations and the
fact that they close only on-shell one introduces external fields and
lets the action depend upon them even bilinearly:
\begin{align}
\Gamma &= \Gcl + \Gamma_{\rm g.f.} + \Gamma_{\rm ext.f.} + \Gamma_{\rm
  bil} \label{Gammaclass}\\
\Gamma_{\rm ext.f.} &= \int \left(\tr \Bigl( Y_{A^\mu} s A^\mu + Y_\lambda
  s\lambda + Y_{\bar\lambda} s\bar\lambda + Y_c s c \Bigr) + (Y_\phi
  s\phi + Y_\psi s\psi + c.c.)\right)\\
\Gamma_{\rm bil} &= \int \tfr{1}{2} \tr \left( Y_\lambda \eps +
  Y_{\bar\lambda} \bar\eps \right)^2 + \left( Y_\psi \eps +
  Y_{\bar\psi} \bar \eps \right)^2\,.
\end{align}
The classical vertex functional $\Gamma$ (\ref{Gammaclass}) satisfies
the ST identity 
\begin{equation}
\S(\Gamma) \equiv \sum_\phi \int \fdq{\Gamma}{Y_\phi}
\fdq{\Gamma}{\phi} + \sum_{\phi'} (s\phi') \fdq{\Gamma}{\phi'} =0
\label{STI}
\end{equation}
(here $\phi$ runs over all non-linearly transforming fields, $\phi'$
over all linearly transforming ones).

It has been shown in \cite{White, PW, HKS} that the gauge condition
\begin{equation}
\fdq{\Gamma}{B} = \xi B + \partial A\,,
\end{equation}
the rigid Ward identity
\begin{equation}
W\, \Gamma = \sum_{\text{all fields $f$}} \int \i \, \delta f\,
\fdq{\Gamma}{f} =0\,,
\end{equation}
the translational ghost equation
\begin{equation}
\pdq{\Gamma}{\omega^\nu} = \pdq{\Gamma_{\rm ext}}{\omega^\nu}
\label{transghosteqn}
\end{equation}
and the ST identity (\ref{STI}) determine the vertex functional
uniquely, if the Adler-Bardeen anomaly is absent and one has supplied
suitable normalization conditions.

The aim is now to study the transformation law of the quantum fields
under the above transformations.

\section{Transformations of the quantum fields
  \label{sec:three}}
\setcounter{equation}{0}
In the perturbative context in which we are working operator equations
can be deduced by using the LSZ reduction formalism, hence we go over
to $Z$, the generating functional of general Green functions.
The translational ghost eqn.\ (\ref{transghosteqn}) leads to the simple WI
\begin{equation}
\sum_f \int \partial_\nu f \, \fdq{\Gamma}{f} =0
\end{equation}
(sum over all fields) which expresses translational invariance of
$\Gamma$.
On $Z$ the WI reads
\begin{equation}
\sum_f \int \partial_\nu j_f \fdq{Z}{j_f} =0\,.
\end{equation}
Rendering the transformations local introduces the
energy-momentum-tensor
\begin{equation}
\sum_f \partial_\nu j_f \, \fdq{Z}{j_f} = \left[ \partial^\mu
  T_{\mu\nu}\right] \cdot Z\,.
\end{equation}
For a string $X$ of propagating fields the respective relation for
Green functions is
\begin{equation}
\sum_\phi \left\langle \pr_\nu \delta(x-y) \phi(x) \,
  X_{\check\phi}\right\rangle = \left\langle \pr^\mu T_{\mu\nu}(x) \,
  X \right\rangle
\end{equation}
($\check\phi$ indicates that $\phi$ is missing in the string). LSZ
reduction w.r.t.\ $X$ yields zero on the l.h.s., on the r.h.s.\ the
matrix element defined by $X$, hence conservation of the
energy-momentum operator 
\begin{equation}
0 = \pr^\mu T_{\mu\nu}^\op\,.
\end{equation}
Reducing w.r.t.\ $X_{\check \phi}$, integrating over all of space
$\int \dv$ and the time slice $(x_0-\mu, x_0+\mu)$ brings about
\begin{equation}
\partial_\nu \varphi^\op = \i \, \left[ P_\nu, \varphi \right]^\op,
\label{transltransop}
\end{equation}
i.e.\ the standard transformation of a quantum field operator under
translations ($\varphi=A_\mu$, $\lambda$, $\bar\lambda$, $B$, $c$,$\bar
c$, $\phi$, $\phi^\dagger$,$\psi$, $\psi^\dagger$).

The BRS charge and transformation properties can be derived from a
version of the ST identity
\begin{equation}
\left( j_{\bar c} \fdq{Z}{j_B}+\sum_\varphi j_\varphi \fdq{Z}{Y_\varphi}\right)
\Bigr|_{\eps=\bar\eps=\omega^\nu=0} = \left[ \pr^\mu
  J_\mu^\brs\right]\cdot Z \Bigr|_{\eps=\bar\eps=\omega^\nu=0}
\end{equation}
in which the gauge BRS transformation has been made to a local one.
For Green functions determined by $X$ this eqn.\ takes the form
\begin{equation}
\delta(x-y) \, \langle B(x)\, X_{\check{\bar c}} \rangle + \sum_\varphi
\delta(x-y) \, \left\langle \fdq{\Geff}{Y_\varphi(x)} X_{\check\varphi}
\right\rangle = \langle \pr^\mu J_\mu^\brs(x)\, X \rangle
\end{equation}
(all at $\eps=\bar\eps=\omega=0$).
Like for the translations it follows first that the BRS charge
\begin{equation}
Q^\brs = \int \dv J_0^\brs
\end{equation}
is conserved (reduction w.r.t.\ $X$), then for the fields that they
transform as
\begin{subequations}
\label{BRStransop}
\begin{align}
\i [Q^\brs, \bar c]^\op &= B^\op\\
\i [Q^\brs, \varphi]^\op &= \fdq{\Geff}{\varphi}^\op
\end{align}
\end{subequations}
($\varphi=A_\mu$, $\lambda$, $\bar\lambda$, $B$, $c$, $\phi$, $\phi^\dagger$,$\psi$, $\psi^\dagger$).
In order to derive the susy transformations we assume all parameters
$\eps$, $\bar\eps$, $\omega$ to be local:
\begin{align}
\S_{\rm loc} Z &\equiv \i j_{A\mu}\fdq{Z}{Y_{A\mu}} + j_{A\mu}
\omega^\nu \pr_\nu \fdq{Z}{j_{A\mu}}
-\i j_c \left( 2\i \eps\sigma^\nu \bar\eps \fdq{Z}{j_A^\nu} -\i
  \omega^\nu \pr_\nu \fdq{Z}{j_c}\right) \nonumber\\
&\quad -\i j_{\bar c} \left( \fdq{Z}{j_B} -\i \omega^\nu \pr_\nu
  \fdq{Z}{j_{\bar c}}\right)
+\i j_B \left( 2\i \eps\sigma^\nu\bar\eps \pr_\nu \fdq{Z}{j_{\bar c}}
  -\i \omega^\nu \pr_\nu \fdq{Z}{j_B} \right) 
\nonumber\\ &\quad
-\i \eta_\lambda^\al \fdq{Z}{Y^\al_\lambda} -\i \bar\eta_{\lambda\da}
\fdq{Z}{\bar Y_{\lambda\da}} 
-\i j_{\phi_i} \fdq{Z}{Y_{\phi_i}} -\i
j_{\phi_i}^\dagger\fdq{Z}{Y_{\phi_i}^\dagger} 
\nonumber\\ &\quad
-\i \eta_{\psi_i}^\al \fdq{Z}{Y_{\eta_i}^\al} -\i \bar\eta_{\psi_i\da}
\fdq{Z}{\bar Y_{\psi_i\da}}
-2\i \eps\sigma^\mu\bar\eps \fdq{Z}{\omega^\mu} 
\nonumber \\
& = \left[ \pr^\mu J_\mu^\brs + \pr^\mu \eps^\al K_{\mu\al} + \pr^\mu
\bar\eps_\da \bar K_\mu^\da + \pr^\mu \omega^\nu K_{\mu\nu} \right]
\cdot Z
\label{STIloc}
\end{align}
($\eta_\lambda$, $\bar\eta_\lambda$, $\eta_{\psi_i}$,
$\bar\eta_{\psi_i}$: sources for $\lambda$, $\bar\lambda$, $\psi_i$,
$\bar\psi_i$). 

$J_\mu$ depends on the ghosts $\eps$, $\bar\eps$, $\omega$ and for
constant ghosts integration yields zero on the r.h.s.

For quite a few steps the calculation runs completely parallel to the
abelian case \cite{RSS}, hence we can be very brief.
Differentiating (\ref{STIloc}) w.r.t.\ $\eps^\al(z)$, the local susy
ghost, (analogously for $\bar\eps_\da(z)$) and performing LSZ
reduction we obtain the operator equation
\begin{equation}
0= \fdq{}{\eps^\al(z)} \pr^\mu J_\mu^\op(x) + \i \T \left( \pr^\mu
  J_\mu(x) \fdq{\Geff}{\eps^\al(z)} \right)^\op + \pr^\mu_x\delta(x-z)
  K_{\mu\al}^\op(x)\,.
\end{equation}
Integration over $z$ yields
\begin{equation}
0= \pr^\mu_x \pr_{\eps^\al} J_{\mu}^\op(x) +\i \int \dz \T \left( \pr^\mu
  J_\mu(x) \fdq{\Geff}{\eps^\al(z)} \right)^\op\,,
\label{star}
\end{equation}
i.e.\ a susy current which is not conserved.
Similarly the susy charge defined by
\begin{equation}
Q_\al(t) \equiv -\int \dv \pr_{\eps^\al} J_0^\op(x)
\end{equation}
will depend on $t$. Integrating (\ref{star}) over all of $x$-space
leads to
\begin{equation}
0 = \int\dt \pr_t Q_\al(t) -\i [Q^\brs, \pr_{\eps^\al}\Geff]^\op
\end{equation}
Taking the time integral for asymptotic times $t=\pm \infty$ and
identifying there the charges we can write
\begin{equation}
Q_\al^\out-Q_\al^\in = \i \left[ Q^\brs, \pr_{\eps^\al} \Geff
\right]^\op\,.
\label{diffcharge}
\end{equation}
Since $Q_\al^\out$ develops out of $Q_\al^\in$ via the time evolution
operator $S$, the scattering operator,
\begin{equation}
Q_\al^\out = S Q_\al^\in S^\dagger \,,
\end{equation}
(\ref{diffcharge}) implies
\begin{equation}
[Q_\al^\in, S] = -\i [ Q^\brs, \pr_{\eps^\al} \Geff \cdot S]\,.
\label{QS}
\end{equation}
Here we have used that $Q^\brs$ and $S$ commute.
The interpretation of this result is clear: the charge $Q_\al^\in$ which
may be taken to be the generator of susy transformations on the free
in-states does not commute with the $S$-operator, the reason being the
$\eps$-dependence of $\Geff$. Looking at (\ref{Gammagf}) this arises from the
gauge fixing term.
Matrix elements between physical states however yield a vanishing r.h.s.\
in (\ref{QS}), hence there $Q_\al^\in$ is a conserved charge.

The transformation law for (non-linearly transforming) fields $\varphi$
can be found by differentiating (\ref{STIloc}) w.r.t.\ $\eps$ and
$\varphi$. After LSZ reduction one obtains the operator relation
\begin{multline}
-\delta(y-x) \fdq{^2\Geff}{Y(y)\delta\eps^\al(z)} -\i \delta(y-x) \T
\left( \fdq{\Geff}{Y(y)} \fdq{\Geff}{\eps^\al(z)}\right) 
\\
= \i \T
\left( \fdq{}{\eps^\al(z)} \pr^\mu J_\mu(y) \varphi(x) \right) -\T \left(
  \pr^\mu J_\mu(y) \fdq{\Geff}{\eps^\al(z)} \varphi(x) \right) 
+ \i \pr_y^\mu \delta(y-x) \T ( K_{\mu\al}(y) \varphi(x) )
\label{SUSYloc}
\end{multline}
Since in the T-product $\T(\pr J \cdot \pr_\eps\Geff \cdot \varphi)$
distributional singularities may arise for coinciding points one
cannot straightforwardly integrate (\ref{SUSYloc}) but has to
determine these singularities. As a consequence of the renormalization
scheme they are well-defined but have to be identified. 
This analysis parallels first again the abelian case: contributions
proportional to $\delta(y-x)$ arise and are cancelled with the
operator product $\T(\delta\Geff/\delta Y(y) \cdot
\delta\Geff/\delta\eps^\al(z))$ on the l.h.s.
But then occurs a change: the contributions which lead to a double
delta function $\delta(x-y)\delta(y-z)$ are represented by diagrams
\begin{equation}
\includegraphics{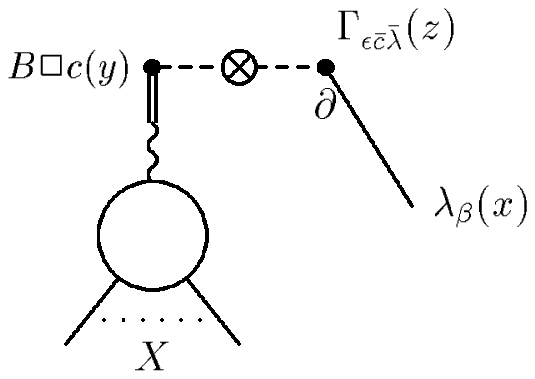}
\label{diagram}
\end{equation}
which differ from SQED in that the ghost propagator $\Delta_{c\bar c}$
is no longer a free one and in that the vertex function $\Gamma_{\eps
  \bar c \bar\lambda}$ receives loop corrections because the ghost
equation of motion is non-trivial. This implies that the local
contribution $a \eps B$ to the transformation $s \lambda$ becomes
order dependent through a coefficient $a$.
A way out of this difficulty is to impose as normalization condition 
\begin{equation}
\left( p^2 \Delta_{c\bar c}\right) \left( \pr_p \Gamma_{\eps\bar c
    \bar\lambda} \right) \Bigr|_{p=0, s=1} =1\,.
\label{normcond}
\end{equation}
As a consequence of the ghost equation 
\begin{equation}
\fdq{\Gamma}{\bar c} + \pr^\mu \fdq{\Gamma}{Y_A^\mu} +\i \omega^\nu
\pr_\nu \fdq{\Gamma}{B} + (2\i\eps\sigma^\nu\bar \eps \pr_\nu \bar c
-\i \omega^\nu \pr_\nu B) \xi =0\,,
\end{equation}
one finds
\begin{align}
\Gamma_{\eps\bar c \bar\lambda} &= -\pr^\mu \pdq{}{\eps}
\fdq{}{Y_A^\mu} \fdq{\Gamma}{\bar\lambda} \\
&= -\pr^\mu \left( \left[ \pdq{}{\eps}\fdq{}{Y_A^\mu} \Geff
  \right]\cdot \Gamma \right)_{\bar\lambda} - \pr^\mu \left( \left[
    \fdq{\Geff}{Y_A^\mu} \right] \cdot \left[ \pr_\eps
    \Geff\right]\cdot\Gamma\right)_{\bar\lambda} \,.
\end{align}
The first term on the r.h.s.\ consists of purely local contributions
whereas the second (a double insertion) is non-local. Hence one can
rewrite (\ref{normcond}) as
\begin{equation}
\left(\sqrt{z_{c\bar c}}\right)^2 \left( a_2 + \left( \left[
      \fdq{\Geff}{Y_A^\mu}\right]\cdot
      \left[\pdq{\Geff}{\eps}\right]\cdot\Gamma\right)_{\bar\lambda}
      \Bigr|_{p=0,s=1} \right) =1\,,
\end{equation}
i.e.\ one fixes indeed the coefficient $a_2$ of the counterterm $\int \bar c \pr ( \i \eps \sigma\bar\lambda -\i \lambda \sigma \bar
\eps )$ by this condition:
\begin{equation}
a_2 = 1 + o(\hbar)
\end{equation}
From tree approximation and with the help of the
symmetric differential operators (see next section) one reads off that
it belongs to the same invariant as $\i \lambda \dslash
\bar\lambda$, the kinetic term.
Imposing this normalization condition one can attribute the double
delta function contribution of diagram (\ref{diagram}) to a local
transformation
\begin{equation}
s\lambda = \eps B + \dots
\end{equation}
(analogously for $\bar\lambda$).

With this result one arrives at the same conclusions as in the abelian
situation.
The term $\eps B$ in the transformation law of $\lambda$ removes the
mismatch between the transformation as given by the above $Q_\al$
(which has been constructed without contribution of the gauge fixing
term) and the time evolution of $\lambda$ as given by its equation of
motion to which the gauge fixing term, of course, contributes.
Hence we have local and evolution compatible transformation laws for
all fields. The breaking of supersymmetry caused by the gauge fixing
is entirely cast into the time dependence of the susy charge.

\section{The anomalous dimensions within the vector multiplet}
We shall now show that the normalization condition (\ref{normcond})
has an interesting consequence for the anomalous dimensions within the
vector multiplet.

The renormalization group (RG) or Callan-Symanzik (CS) equation which
are identical in the massless case follow as usual \cite{PS} as
differential equations symmetric with respect to the ST identity and
compatible with the gauge condition.
A basis of the respective differential operators is provided by
\begin{align}
\N_A &\equiv \int A \fdq{}{A} - Y_A \fdq{}{Y_A} - B \fdq{}{B} + \bar c
\fdq{}{\bar c} \\
\N_\lambda &\equiv \int \cdq{\lambda} -\cdq{Y_\lambda} + c.c. \\
\N_c &\equiv \int -\cdq{c} + \cdq{Y_c} \\
\N_\phi &\equiv \int\cdq{\phi} - \cdq{Y_\phi} + c.c.\\
\N_\psi &\equiv \int \cdq{\psi} -\cdq{Y_\psi} + c.c.
\end{align}
The CS operator can be expanded in this basis and the CS eqn. then
reads
\begin{equation}
\C \Gamma \equiv \left( m\pr_m + \beta_g\pr_g -\sum_\varphi \gamma_\varphi
  \N_\varphi\right) \Gamma =0
\label{CS}
\end{equation}
($m$ comprises all mass parameters, $g$ all couplings of the theory,
$\varphi$ runs over $A$, $\lambda$, $c$, $\phi$, $\psi$).
This implies that apriori the anomalous dimension $\gamma_A$ of the
vector field is not related to $\gamma_\lambda$, the anomalous
dimension of $\lambda$, $\bar\lambda$.

Testing (\ref{CS}) w.r.t.\ $c$ and $\bar c$ we first find
\begin{equation}
(m \pr_m + \beta_g \pr_g) \Gamma_{c \bar c} = (\gamma_A - \gamma_c)
\Gamma_{c \bar c}\,,
\end{equation}
hence \begin{equation}
(m \pr_m + \beta_g \pr_g) \Delta_{c \bar c} = (\gamma_c -
\gamma_A)\Delta_{c \bar c} \,.
\end{equation}
Similarly,
\begin{equation}
(m \pr_m + \beta_g \pr_g) \Gamma_{\eps \bar c \bar\lambda} = (\gamma_A
-\gamma_\lambda) \Gamma_{\eps \bar c \bar\lambda} =0\,. 
\end{equation}
Hence, acting with $(m\pr_m + \beta_g\pr_g)$ on (\ref{normcond}) we obtain
\begin{equation}
(\gamma_c-\gamma_A) +(\gamma_A-\gamma_\lambda)=0\,,
\end{equation}
i.e.\
\begin{equation}
\gamma_\lambda = \gamma_c\,. \label{gammalambdac}
\end{equation}
The normalization condition (\ref{normcond}) leads to coinciding
anomalous dimensions for the gaugino and ghost fields.
Even more can be said about the anomalous dimensions when we choose
the Landau gauge $\alpha=0$. In this gauge, an additional ghost equation
\cite{MPW} holds,
\begin{align}
\F \Gamma \Bigr|_{{\rm ext.f.}=0} &=0\,,\\
\F &= \int \left( \fdq{}{c} + \i g \left[ \bar c, \fdq{}{B} \right] 
\right) \,.
\end{align}
By acting with the commutation relation
\begin{equation}
\left[ \F, \C \right] = (\gamma_c-2 \gamma_A) \int \fdq{}{c} 
\end{equation}
on $\Gamma$, we obtain
\begin{equation}
\gamma_c=2\gamma_A \,, \label{gammaAc}
\end{equation}
and together with (\ref{gammalambdac})
\begin{equation}
\gamma_\lambda=2\gamma_A\,. \label{gammalambdaA}
\end{equation}
In an arbitrary gauge, (\ref{gammaAc}) may be imposed as a
normalization condition, fixing the counterterm $\S_{\Gamma} \int \tr
(c Y_c)$. 

The result (\ref{gammalambdaA}) fits to the asymptotic supersymmetry which also followed
as a consequence of it as shown in the last section. It is not in
contradiction to previous calculations \cite{ATV} because there other
normalization conditions have been used and anomalous dimensions
depend quite generally on normalization conditions and gauge fixing.

Without further inquiries nothing can be said about the anomalous
dimensions within the matter multiplets. It may however very well turn
out that a suitable non-linear gauge fixing could render them equal as
well. That would be of importance in models with $N=2$ symmetry.
For the $N=4$ theory, too, the role of gauge fixing and its effect on
anomalous dimensions seems not to have been discussed thoroughly.

\section{Discussion and conclusions}
In the context of $N=1$ SYM theory we studied the transformation
properties of quantum fields in perturbation theories. The generalized
ST identity (\ref{STIloc}) with local ghosts for translations, BRS and susy
transformations governs these transformation laws via LSZ
reduction. As already remarked in \cite{RSS} this implies
representation dependence of operator equations whose validity outside
of perturbation theory is not clear.

For translations and BRS we found the standard
transformations (\ref{transltransop}), (\ref{BRStransop}) which is not
surprising since both are conserved.
Since the gauge fixing breaks susy the respective charge is not
conserved in Fock space but only in the Hilbert space of the
theory. It turns out that the vectorino fields $\lambda$,
$\bar\lambda$ receive additional contributions in their transformation
law. They can be understood as proper, local transformations $\eps B$,
$\bar\eps B$ if one imposes a suitable normalization condition
(\ref{normcond}) which determines the relevant part of gauge fixing.
For the fields $\lambda-\eps \bar c$, $\bar\lambda-\bar\eps \bar c$
one then has local susy transformations such that asymptotically susy
is restored. This normalization condition also implies that the
anomalous dimension of $\lambda$ equals that of the ghost, and in the
Landau gauge (or by imposing an additional normalization condition) it
is related to the anomalous dimension of the vector field $A_\mu$.

A comment is in order as to the validity of our derivation. We have
written all formulae for a completely massless theory. Then, of
course, there is no S-matrix and one cannot go on-shell.
Hence we may add mass terms for all fields and thus maintain in
general only translational invariance but neither ordinary BRS nor
susy. Our equations should then be read modulo soft breaking terms --
which is sufficient for clarifying the role of gauge fixing and
normalization condition (\ref{normcond}). For models where susy is
maintained but the gauge invariance is completely broken down such
that all vector fields (and the vectorinos) get mass our results
should be strictly true.
Alternatively one could have derived within the massless theory the
desired operator equations by performing the reduction not with
respect to the asymptotic states but with suitable other combinations.

\end{document}